\documentclass[journal]{IEEEtran}
\usepackage{ifpdf}
\usepackage{cite}
\usepackage[pdftex]{graphicx}
\usepackage[cmex10]{amsmath}
\usepackage{array}
\usepackage{mdwmath}
\usepackage{mdwtab}
\usepackage{eqparbox}
\usepackage{amssymb}
\usepackage[tight,footnotesize]{subfigure}
\usepackage{url}
\usepackage{textcomp}
\usepackage[hyperindex]{hyperref}
\usepackage{amssymb}     
\usepackage{marvosym}    
\usepackage{upgreek}     
\usepackage[none]{hyphenat}
\usepackage[font=footnotesize]{caption}
\usepackage[justification=centering]{caption} 
\newcommand{\gammas}      {$\gamma${s}}

\newcommand{\onbb}        {{$0\nu\beta\beta$}}

\newcommand{\gesix}       {{$^{76}$Ge}}
\newcommand{\geenr}       {{$^{\rm enr}$Ge}}          
\newcommand{\genat}       {{$^{\rm nat}$Ge}}

\begin{document}
\title{Status Report of the \textsc{Gerda} Phase II Startup}
\author{Valerio D'Andrea, on behalf of the \textsc{Gerda} collaboration
\thanks{V. D'Andrea  is with the Gran Sasso Science Institute (INFN), L'Aquila, Italy
(email: \href{mailto: valerio.dandrea@gssi.infn.it}{valerio.dandrea@gssi.infn.it}).} }
\maketitle
\thispagestyle{empty}

\begin{abstract}
 The GERmanium Detector Array (\textsc{Gerda}) experiment, located at the Laboratori Nazionali del Gran Sasso (LNGS) of INFN, searches for $0\nu\beta\beta$ of $^{76}$Ge. Germanium diodes enriched to $\sim 86~\%$ in the double beta emitter
\gesix (\geenr) are exposed being both source and detectors of $0\nu\beta\beta$ decay.
Neutrinoless double beta decay is considered a powerful probe to address still open issues in the neutrino sector of the (beyond) Standard Model of particle Physics.

Since 2013, just after the completion of the first part of its experimental program (Phase I), the \textsc{Gerda} setup has been upgraded to perform its next step in the $0\nu\beta\beta$ searches (Phase II). Phase II aims to reach a sensitivity to the $0\nu\beta\beta$ decay half-life larger than $10^{26}~$yr in about 3 years of physics data taking.
This exposing a detector mass of about $35~$kg of \geenr~and 
with a background index of about $10^{-3}~$cts/(keV$\cdot$kg$\cdot$yr).

One of the main new implementations is the liquid argon scintillation light read-out, to veto those events that only partially deposit their energy both in Ge and in the surrounding LAr.

In this paper the \textsc{Gerda} Phase II expected goals, the upgrade work and few selected features from the 2015 commissioning and 2016 calibration runs will be presented.  The main Phase I achievements will be also reviewed.
\end{abstract}


\IEEEpeerreviewmaketitle

\section{Introduction}
\IEEEPARstart{T}{he} GERmanium Detector Array (\textsc{Gerda}) \cite{gerda_ep-2013} is an experiment searching for the neutrinoless double beta ($0\nu\beta\beta$) decay of $^{76}$Ge.
High-purity germanium (HPGe) detectors enriched in $^{76}$Ge to $\sim 86\%$ are exposed acting both as sources and detection media.

The experimental signature of $0\nu\beta\beta$ decay is a peak in the spectrum of the summed energies of the two electrons, released in the nuclear process. The peak is expected at the $Q_{\beta\beta}$ value that for $^{76}$Ge is $2039~$keV.

The events from $0\nu\beta\beta$ decays can be distinguished from $\gamma$-induced background; infact it is expected that the energy deposition of the two emitted electrons is localized within few mm (single-site events, SSE) while $\gamma$-rays deposit energy in several positions (multi-site events, MSE).

The signals are different for SSE and MSE; surface events from $\alpha$ or $\beta$ decays also exhibit characteristic shapes. Thus, pulse shape discrimination (PSD) techniques can improve the experimental sensitivity.

The half-life on $0\nu\beta\beta$ decay can be calculated as \cite{sensitivity}:
\begin{equation}
T^{0\nu}_{1/2} = \frac{\ln 2 \cdot N_A}{m_{enr}\cdot N^{0\nu}} \cdot M \cdot t\cdot \varepsilon
\end{equation}
where $N_A$ is the Avogadro number, $m_{enr}$ is the isotopic mass of the considered isotope, $N^{0\nu}$ is the observed signal strength, $M$ the total source mass and $t$ the live time of the measurement.
The total efficiency $\varepsilon$ is the product of several factors:
\begin{equation}
 \varepsilon = f_{76} \cdot f_{av} \cdot \varepsilon_{FEP} \cdot \varepsilon_{PSD}
\end{equation}
$f_{76}$ is the fraction of $^{76}$Ge atoms, $f_{av}$ is the active volme fraction, $\varepsilon_{PSD}$ is the signal acceptance by the PSD and $\varepsilon_{FEP}\sim90\%$ is the probability that a $0\nu\beta\beta$ decay releases its entire energy in the detector, losses are due to bremsstrahlung photons, fluorescence X-rays or electrons escaping the detector active volume.

\section{The \textsc{Gerda} Experiment}
\begin{figure}[!t]
\centering
\includegraphics[width=0.43\textwidth]{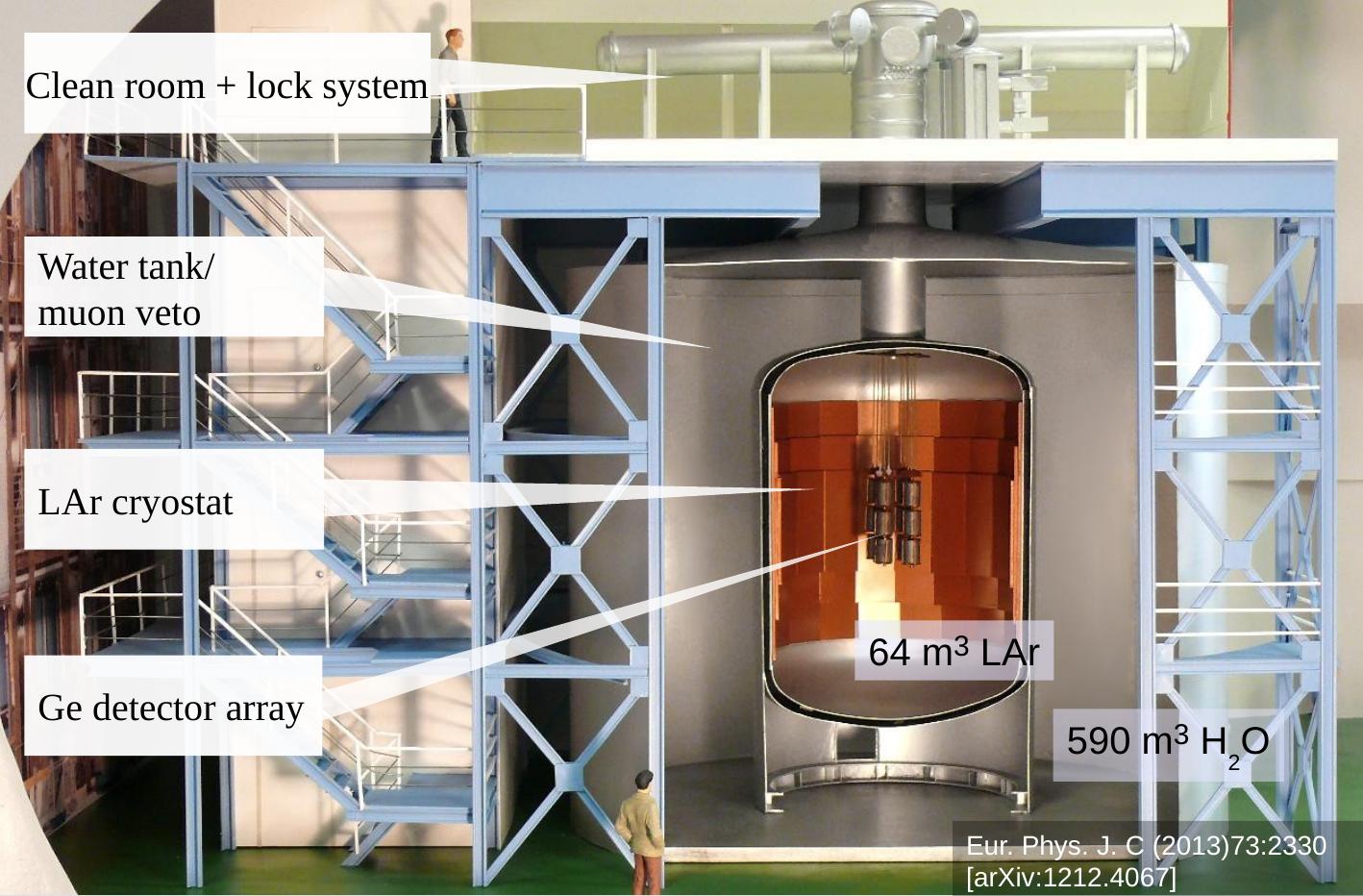}
\caption{Section of the \textsc{Gerda} experiment}
\label{gerda-view}
\end{figure}

The construction of \textsc{Gerda} was tailored to background minimization. The germanium detectors are mounted in low mass ultra-pure holders and are directly inserted in $64~$m$^3$ of liquid argon (LAr), acting as cooling medium and shield against external background radiation (Fig. \ref{gerda-view} shows the section of the \textsc{Gerda} setup).
The argon cryostat is complemented by a water tank with $10~$m diameter which further shields from neutron and $\gamma$ backgrounds. It is instrumented with photomultipliers to veto the cosmic muons by detecting \v{C}erenkov radiation. The muon veto hermeticity is provided by plastic scintillators installed on the top of the structure. A detailed description of the experimental setup is in Ref. \cite{gerda_ep-2013}.

A first physics data collection, denoted as Phase I, was carried out from November 2011 to June 2013. In this phase eight p-type semi-coaxial detectors enriched in $^{76}$Ge from the Heidelberg-Moscow (HdM) \cite{hdm} and IGEX \cite{igex} experiments and five Broad Energy Germanium (BEGe) detectors were used \cite{bege}. Three coaxial detectors with natural isotopic abundance from the Genius Test Facility (GTF) project \cite{gtf,gtf2} were also installed.

In 2013, at the completion of Phase I, the setup upgrade was started; in 2015 many test and commissioning runs were performed and since January 2016 the Phase II physics data taking is ongoing.

\subsection{Phase I Results}
From November 2011 to May 2013 a global exposure of $21.6~$kg$\cdot$yr has been collected. Data blinding was implemented for the first time in a $0\nu\beta\beta$ experiment data analysis protocol. Events with energies within $Q_{\beta\beta}\pm20~$keV were not processed.
After the fine energy calibration, the PSD criteria and the background model were finalized the window was opened except for $\pm5~$keV ($\pm4~$keV) around $Q_{\beta\beta}$ for the semi-coaxial (BEGe) detectors; only at the end the $Q_{\beta\beta}$ region were analyzed.

The \textsc{Gerda} background model \cite{bkg} predicts a flat energy distribution between 1930 and 2190 keV from Compton events of $\gamma$ rays of $^{208}$Tl and $^{214}$Bi decays, degraded $\alpha$ events and $\beta$ rays from $^{42}$K and $^{214}$Bi.

In the range $Q_{\beta\beta} \pm 5~$keV seven events are observed before the PSD, to be compared to $5.1 \pm 0.5$ expected background counts. Three out of the six events from the semi-coaxial detectors are classified as SSE by the artificial neural network (ANN) method, consistent with the expectation.
The event in the BEGe data set is rejected by the $A/E$ cut (the powerful PSD method for the BEGes that uses the ratio between the ampitude of the current pulse $A$ and the total energy $E$), hence no events remain within $Q_{\beta\beta} \pm \sigma_E$ after PSD. 

\begin{figure*}[!t]
\centering
\includegraphics[width=0.5\textwidth]{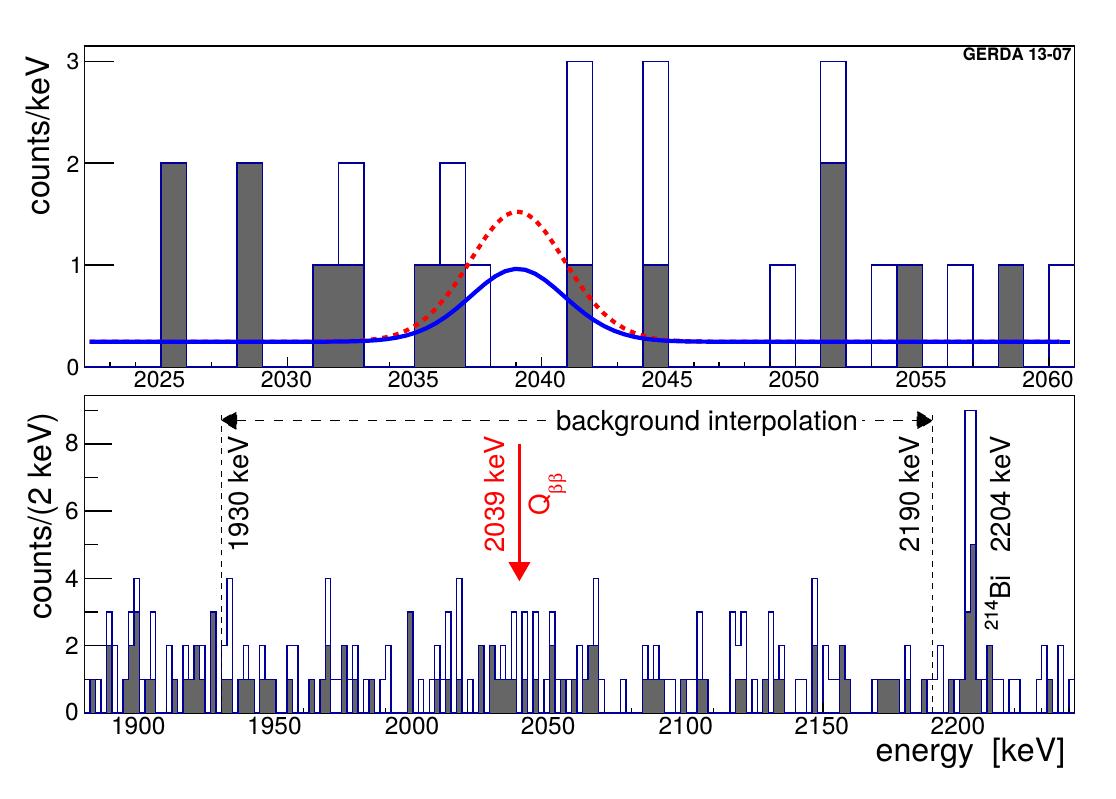}
\caption{Combined energy spectrum from all \geenr~detectors without (with) PSD is shown by open (filled) histogram \cite{gerdaI-res}}
\label{spectrum-psd}
\end{figure*}

The combined energy spectrum around $Q_{\beta\beta}$, with and without the PSD selection, is displayed in Fig. \ref{spectrum-psd}.

To derive the signal strength $N_{0\nu}$ and a frequentist coverage interval, a profile likelihood fit was performed \cite{gerdaI-res}. The fitted function consists of a constant term for the background and a Gaussian peak for the signal with mean at $Q_{\beta\beta}$ and standard deviation $\sigma_E$ according to the expected resolution.

The best fit value is $N_{0\nu} = 0$ that means no signal events above the background.
The derived half-life limit on $0\nu\beta\beta$ decay is:
\begin{equation}
 T_{1/2}^{0\nu} > 2.1 \cdot 10^{25} \text{yr} ~ \text{(90\% C.L.)}
\end{equation}
including the systematic uncertainty.

\textsc{Gerda} Phase I data show no indication of a peak at $Q_{\beta\beta}$ and the claim for the observation of $0\nu\beta\beta$ \cite{klapdor} decay in $^{76}$Ge is not supported.

This result is consistent with the limits by HdM and IGEX experiments and extending the profile likelihood fit including the data sets of those experiment the best fit yields is still $N_{0\nu} = 0$ and a new limit for the $0\nu\beta\beta$ decay is established:
\begin{equation}
 T_{1/2}^{0\nu} > 3.0\cdot 10^{25} \text{yr} ~ \text{(90\% C.L.)}
\end{equation}

In addition to the new limit on the $0\nu\beta\beta$, the Phase I data allowed to establish the new value for the two neutrino accompanied $\beta\beta$ decay ($T_{1/2}^{2\nu}=1.926\pm0.094 \cdot 10^{21}~$yr), to search for neutrinoless $\beta\beta$ decay processes accompanied with Majoron emission (no signals were found and lower limits of the order of $10^{23}~$yr were set) \cite{2nubb-majoron} and to study the two neutrino $\beta\beta$ decay of $^{76}$Ge to excited states of $^{76}$Se \cite{2nubb_excited} (no signal observed and new bounds determined for three transition, two orders of magnitude larger than those reported previously).

\section{Upgrade to Phase II}
The \textsc{Gerda} Phase II goal is the reduction of a factor of ten the background obtained in Phase I and this can only be achieved with an optimized experimental design.
After several years of R \& D, a version of the broad energy germanium (BEGe) detector \cite{canberra} from Canberra with a thick entrance window has been selected.
The advantages of using the BEGe detectors are their superior rejection of background with a simple and powerful analysis and their optimal energy resolution due to a very low detector capacitance.

In addition an active suppression of background by detecting the LAr scintillation light is introduced.

\begin{figure}[!t]
\centering
\includegraphics[width=2.in]{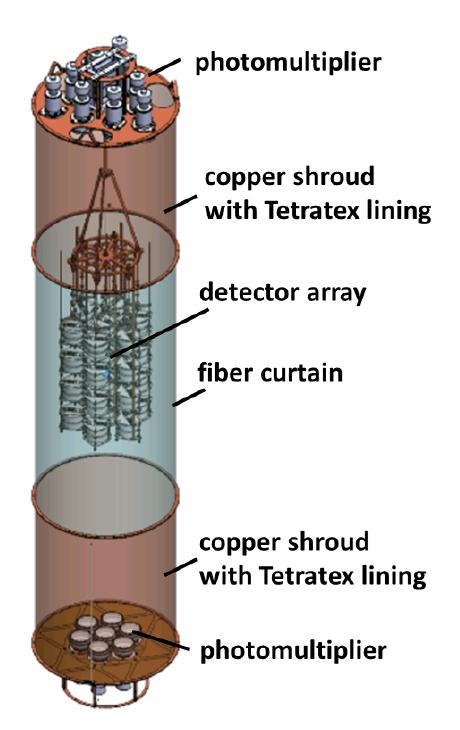}
\caption{Phase II assembly of detector array and LAr veto system as it will be immersed into the cryostat}
\label{array2}
\end{figure}

Fig. \ref{array2} shows the core of the Phase II \textsc{Gerda} setup: the Ge detector array, whose mass is doubled compared to Phase I, is at the center of a vetoed LAr volume.

The design allows to assemble both the detector array and the surrounding LAr veto system in the open lock under dry nitrogen atmosphere and to lower both systems together into the cryostat.

\subsection{BEGe Detectors}
\begin{figure}[!t]
\centering
\includegraphics[width=2.2in]{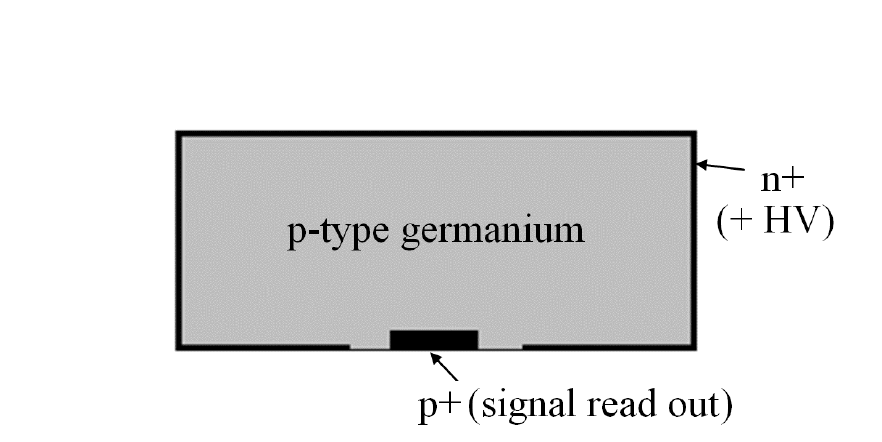}
\caption{Schematic drawing of a BEGe detector}
\label{bege}
\end{figure}
The BEGe detectors for Phase II (5 of them already used in Phase I) are a modified model BE5030 available from Canberra Semiconductor (Olen). A schematic view of a BEGe diode is shown in Fig. \ref{bege}.

The detector is made of p-type HPGe with the Li$^-$ drifted n$^+$ contact ($0.7~$mm specified thickness) covering the whole outer surface, including most of the bottom part. The small p$^+$ contact is located in the middle of the bottom side.

The raw material for the BEGe diodes enriched in \gesix has been produced in form of $53.3~$kg of $^{\rm{enr}}$GeO$_2$ from ECP (Zelonogorsk, Russia) with the \gesix isotope enriched to $\sim 88~\%$. The reduction and purification of the GeO$_2$ was achieved with an efficiency of $94\%$ yielding $35.5~$kg $^{\rm enr}$Ge(6N) for crystal production at Canberra (Oak Ridge), a total of 9 crystals could be pulled.
The crystals were cut into 30 slices and sent to Canberra (Olen) where they were transformed into working BEGe detectors with a total mass of $20.0~$kg.

All detectors have been characterized in the HADES underground facility close to Olen; the relevant operational parameters including active volumes, dead layers and pulse shape performances have been defined. 29 out of 30 detectors work according to specifications reaching full depletion with bias voltages below $5~$kV and an energy resolution at $1.3~$MeV of $< 1.9~$keV of FWHM when operated in a standard vacuum cryostat and standard Canberra Ge detector front end electronics \cite{bege}.
In a final step, Al pads for wire bonding were evaporated on the p$^+$ and n$^+$ substrates of each crystal.

During all production steps, the exposure of the enriched material to the cosmic radiation has been reduced significantly by shielded transport and underground storage of the material.

The two main adavantages of using these detectors are the optimal energy resolution due to the very low input capacitance ($\sim~$pF) and the powerful pulse shape discrimination. The particular position of the p$^+$ and n$^+$ contacts produces a non-uniform electrical field inside the BEGe detectors, allowing to discriminate the single-site (like the \onbb) from the multi-site events using the $A/E$ ratio: the SSE events have a fixed $A/E$ and can be discriminate from the typical background events normally MSE.

\subsection{Detector Assembly and Read-Out Electronics}
To provide a further background reduction a new holder for detectors, new contacts and new read-out electronics were implemented for Phase II.

Since the model of the \textsc{Gerda} Phase I background \cite{bkg} showed that a large fraction of the background is originated from sources close to the Ge detectors, in Phase II most of the material close to the Ge detector array were replaced by material of higher radio-purity and reduced mass.
The Phase I spring-loaded detector contacts have been replaced by wire bonded ones.
Moreover the new holder is made by a plate of $40~$g of mono-crystalline silicon which is intrinsically radio-pure. The silicon plate provides also the fixation of both the signal and the HV contacts.

The test and commissioning runs showed that, for different reasons,  many detectors had reverse current. Many detectors were sent to Canberra for groove passivation and the back-to-back detector mounting assemblies were modified to have the groove looking downward. This to prevent particles floating in the LAr depositing and stay in the groove triggering the reverse current.
As a consequence of the taken actions the fraction of detectors showing reverse current significanty decreased.

To take full advantage of the low input capacitance of the BEGe diodes and to accomplish the more stringent requirements in terms of radio-purity of the Phase II also the front-end electronics was improved. A new version of the charge sensitive preamplifiers, called CC3, was developed for \textsc{Gerda} Phase II \cite{cc3}.

\begin{figure}[!t]
\centering
\includegraphics[width=1.8in]{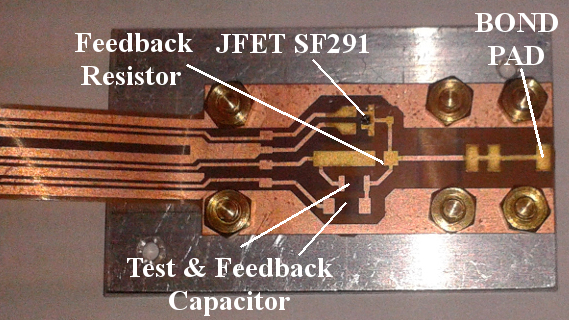}
\caption{Very Front End section of CC3 preamplifier designed for \textsc{Gerda} Phase II}
\label{vfe}
\end{figure}

The new preamplifier CC3 is composed of two separate sections: the Very-Front-End (VFE) section (Fig. \ref{vfe}), consisting of the preamplifier input JFET, the feedback components (resistor and capacitor) and the main section of the preamplifier, consisting of operational amplifiers, passive components and connectors.

The VFE section is mounted very close to the detector to improve its performances. For this reason it must be extremely radio-pure: the input JFET is a commercial bare die (SF291) connected with bonding wire, the discrete feedback resistors (of $500~$M$\Omega$) is a 0402 SMD ceramic package and the feedback capacitor ($0.3~$pF) is a printed trace.
The flexible VFE circuit has a coplanar waveguide tail to transmit the three relevant signal (test, feedback and JFET drain) from/to the CC3.

The high mortality of the bare die JFETs during the detector handling operations motivated to keep the Phase I readout scheme, with the BF862 JFET close to the second amplifying stage located at $40~$cm ($80~$cm) above the top (bottom) detector in the string.
The detector-CSA contacts, the CSA boards and the full coaxial cabling (for Ge, PMTs, SiPMs) have been redesigned and made out of new material with respect to Phase I; as result their $^{232}$Th, $^{238}$U, $^{226}$Ra contents is reduced by a factor of 3 for the front-end devices and factor 10 for unit length of the coaxial cables.

The performances obtained by the coupling of the CC3 preamplifier and BEGe detectors is a FWHM of $\sim 2.6~$keV at 2.6 MeV, whit a $20~$MHz bandwidth allowing PSD by the A/E method, $50~$mW/channel power dissipation that is suitable for operation in LAr and a radio-purity of $50~\mu$Bq/PCB for $^{228}$Th including connectors.

\subsection{LAr Instrumentation}
Fig. \ref{array2} shows the LAr veto system: 9 top and 7 bottom PMTs \cite{pmt} collect light from a LAr volume of $220~$cm height and 49 diameter surrounding the Ge detector array. A curtain of wavelength shifting fibers, coupled to silicon photomultipliers (SiPMs) \cite{sipm}, encloses the middle $100~$cm length of this volume and can collect light also outside the diameter of the enclosed cylinder. Hence the diameter of the vetoed LAr volume at the Ge detector array level is enlarged.

The upper and lower parts of the vetoed volume, $60~$cm in height each, consist of thin-walled ($0.1~$mm) copper cylinders lined by wavelength shifting reflector foils (Tetratex + TPB), while the fiber curtain is made of $1\times1~$mm wavelength shifting fibers coated with TPB and coupled in groups of nine to $3\times3$ SiPM arrays.

Pointing with their diagonals to the central axis, the total of 810 fibers covers almost $80\%$ of the circumference.

\section{Start of Phase II Data Taking}
The commissioning phase started in July 2015 and several runs were performed with different detector configurations.

On December 20th, 2015 the Phase II configuration with all the detectors mounted in the array was achieved and
the data taking has been started.

\begin{figure}[!h]
\centering
\includegraphics[width=0.4\textwidth]{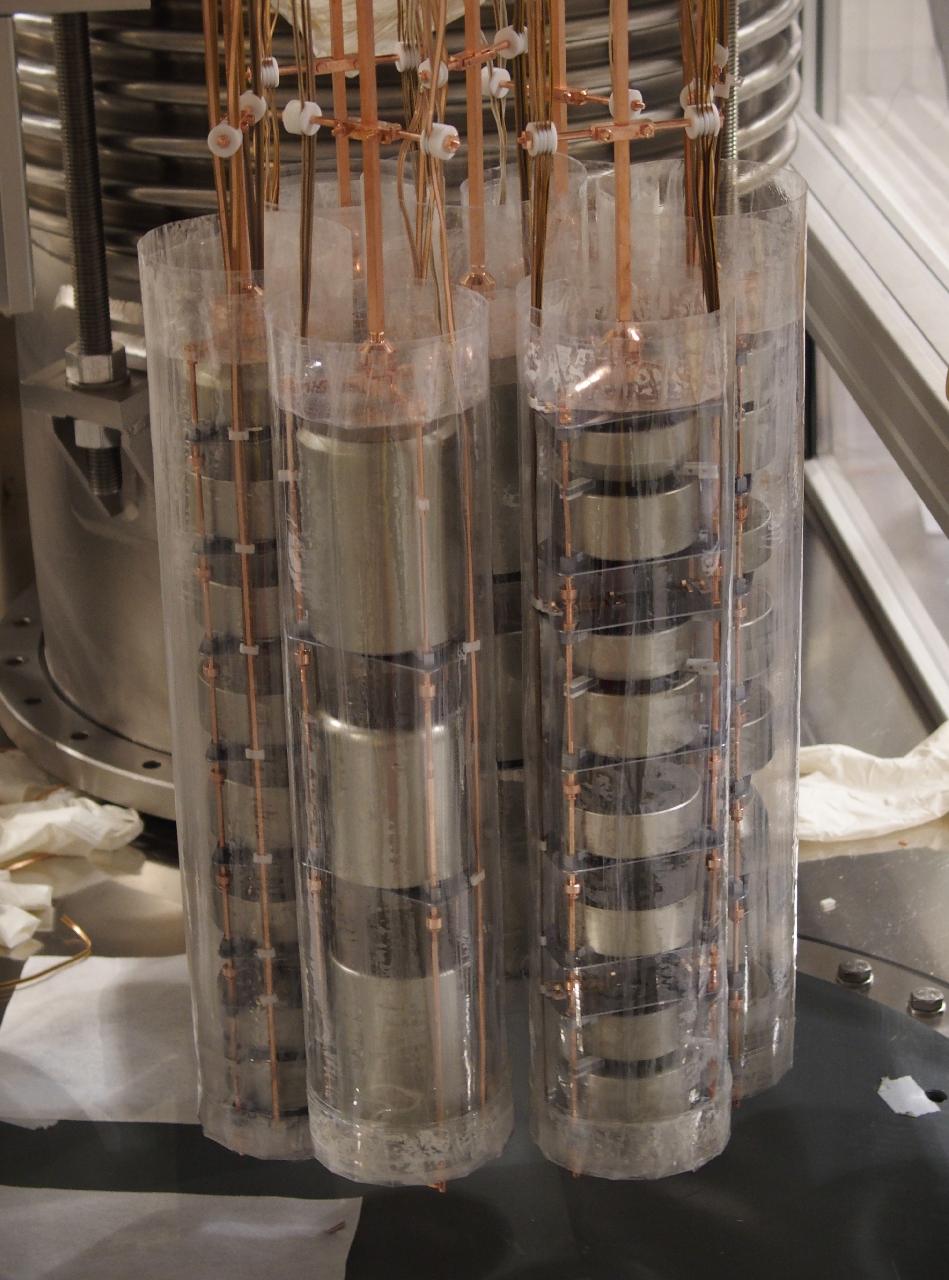}
\caption{Full detector array of \textsc{Gerda} Phase II mounted on December 2015}
\label{strings}
\end{figure}

\begin{figure}[!t]
\centering
\includegraphics[width=0.50\textwidth]{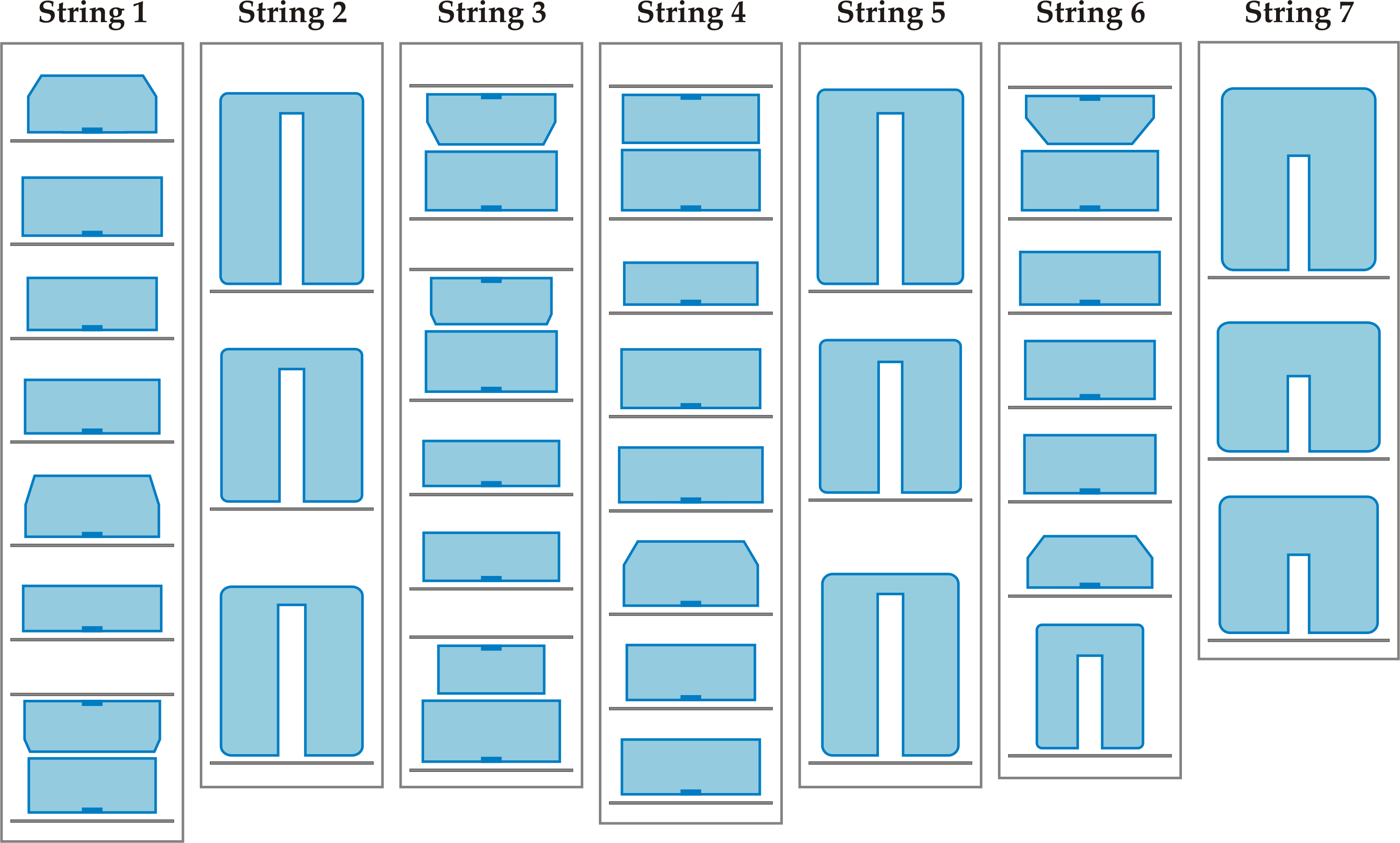}
\caption{Schematic view of the strings: the BEGe detectors are arranged in 4 strings (in string 6 there is also a coaxial), the coaxial in the other 3 strings}
\label{strings_scheme}
\end{figure}

Fig. \ref{strings} shows a picture of the full array just before the immersion in LAr: seven \geenr~ plus three natural coaxials (both from Phase I) and thirty new \geenr~ BEGe detectors for a total of 40 detectors accounting for $35.6~$kg of \geenr~ and $7.6~$kg of \genat~ are organized in seven strings; Fig. \ref{strings_scheme}  shows a schematic view of the the detector arrangement.
The BEGes ($20~$kg of \geenr) are organized in three strings of eight and one string of six detectors, the coaxials ($15.6~$kg of \geenr) are organized in three strings of three detectors, plus one coaxial in a BEGe string.
Each detector string is surronded by a nylon mini-shroud, preventing the $^{42}$K ions from being drifted and diffused at the detector surfaces. 

After the immersion of the array all 40 detectors are working, most of them are at operational voltage showing a leakage current $<100~$pA, 3 BEGes and one coaxial are showing a value of LC$~\gtrsim 100~$pA and needed a decrease of the bias voltage.

The energy scale of detectors is determined by weekly irradiating the Ge detector array by three $^{228}$Th ($20$-$30~$kBq) sources. The stability of the setup is monitored comparing the reconstructed peak positions in different calibrations.

\begin{figure}[!t]
 \centering
 \includegraphics[width=0.50\textwidth]{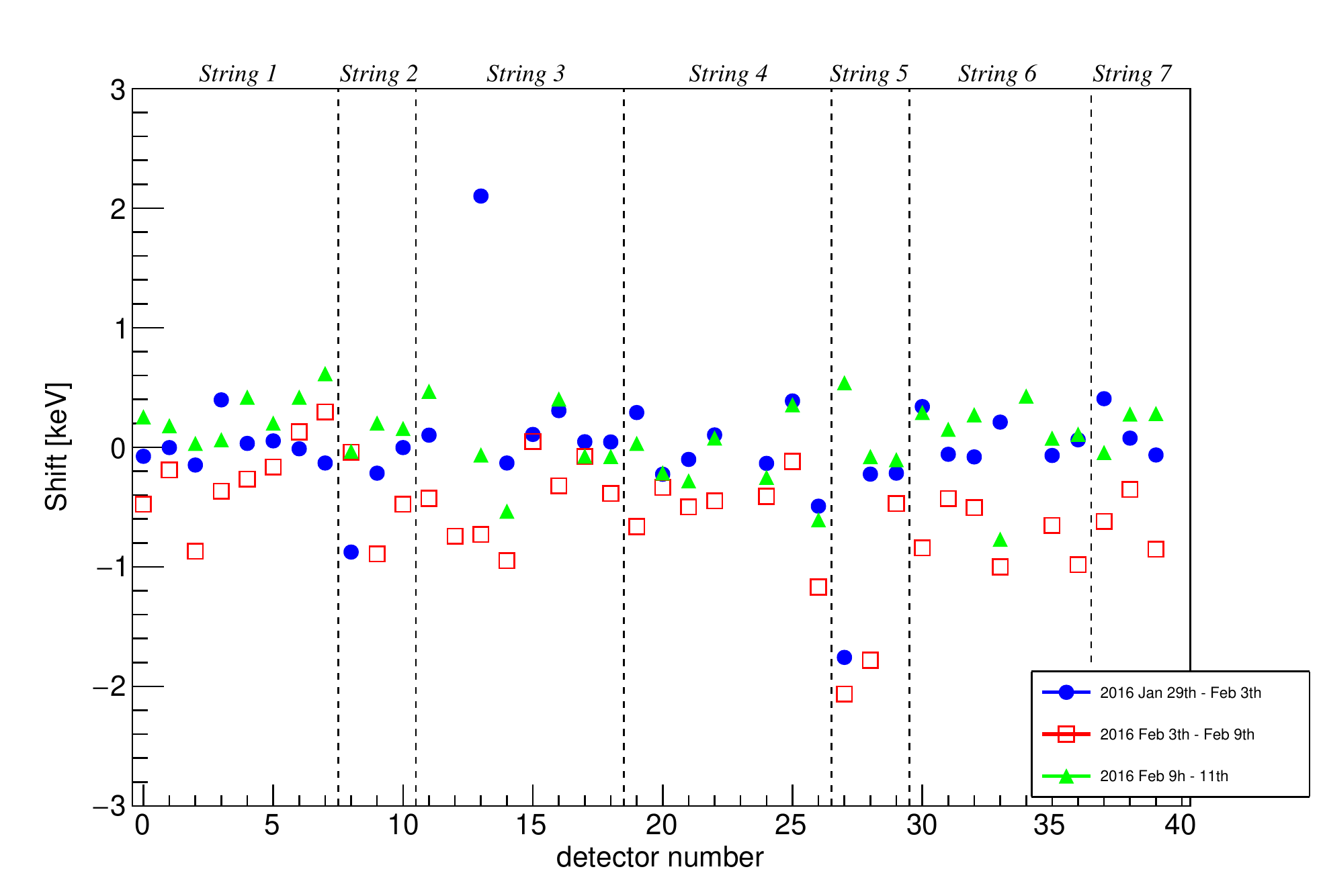}
 \caption{Energy shift of the $2614.5~$keV peak in the comparison between the calibration runs of January 29th and February 3rd (in blue), February 3rd and February 9th (in red) and February 9rd and February 11th (in green). The horizontal axis follows the detector progressive number and the top legend reports the string number}
 \label{shift}
\end{figure}

Fig. \ref{shift} shows the energy shift for the $2614.5~$keV peak between the calibration runs of January 29th, February 3rd, February 9th and February 11th.

The horizontal axis reports the detector progressive number following its arrangement in the strings starting from string 1 to string 7. Excluding few cases, the shifts are in the range between $-1~$keV and $+1~$keV.

The performances of the setup are also established using the calibration data. The energy spectrum is reconstructed and the resolution (in terms of FWHM) of the peaks is calculated.
In \textsc{Gerda} a novel filtering method was developed in order to improve the energy resolution \cite{zac}.

\begin{figure}[!t]
 \centering
 \includegraphics[width=0.5\textwidth]{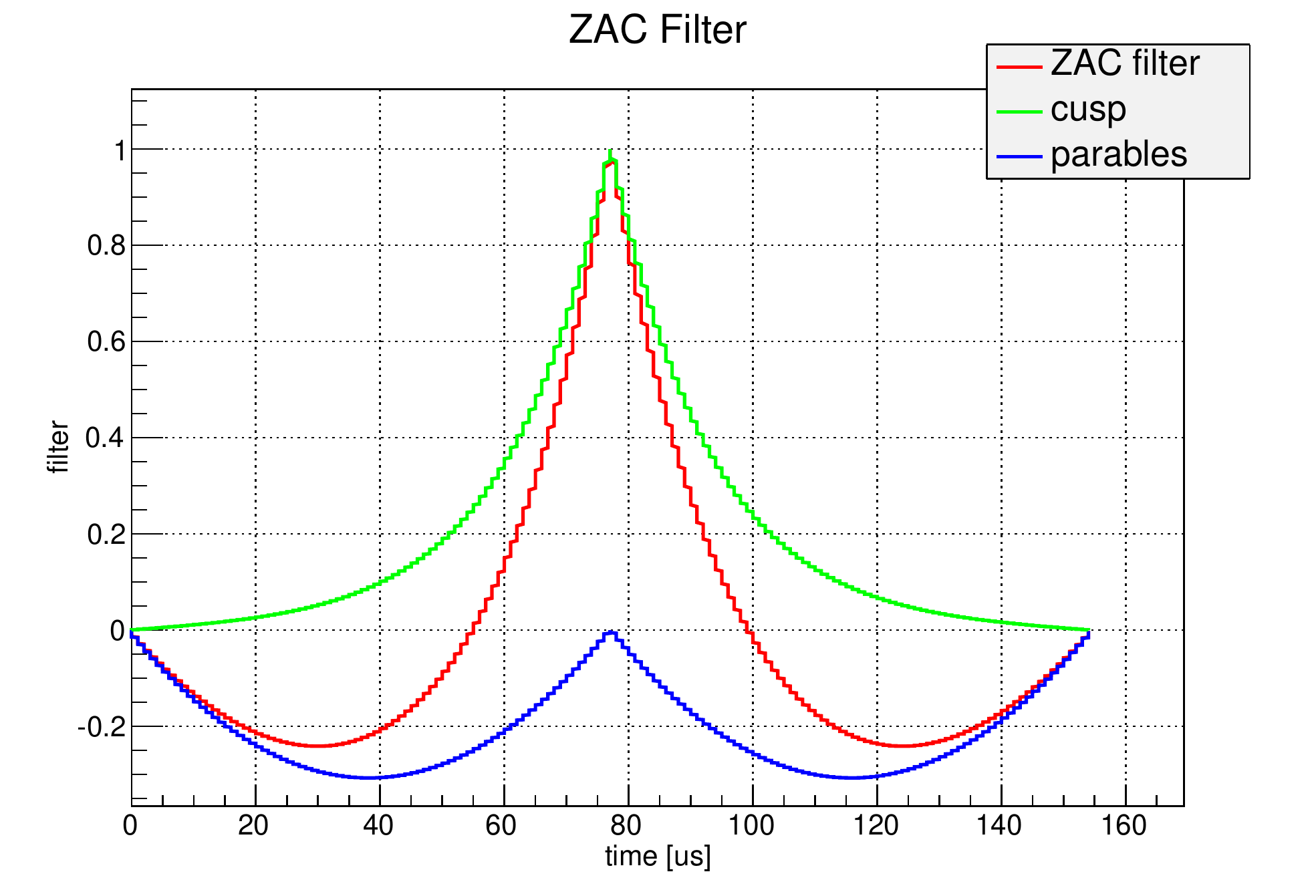}
 \caption{Zero-Area finite-length Cusp (ZAC) filter (in red), composed of the finite-length cusp (in green) from which two parabolas are subtracted on the cusp sides (in  blue)}
 \label{zac_filter}
\end{figure}

From noise theory \cite{Gatti-Manfredi}, it can be shown that the infinite cusp is the optimal filter to abate both series and parallel noise. In a real case where a limited waveform length is recorded ($160~\mu$s in \textsc{Gerda}) and in presence of low frequency disturbances, a Zero Area, modified Cusp (ZAC), finite impulse response filter has shown to improve the energy resolution of about $0.3~$keV both for coaxials and BEGes, whit respect to gaussian filter. An example of such filter is in Fig. \ref{zac_filter}, the central flat-top (of $\sim 1~\mu$s) allows for full charge collection.

Applying this optimized method to the Phase I calibration data, the average energy resolutions at the $^{208}$Tl $2614.5~$keV line were of $(2.76^{+0.08}_{-0.06})~$keV and $(4.32^{+0.38}_{-0.30})~$keV for BEGes and coaxials respectively \cite{zac}. The averages are evaluated among different detectors on tenths of independent calibrations. The errors represents the FWHM half-spread among 4 (6) BEGes (COAX) detectors.
 
\begin{figure}[!t]
 \centering
 \includegraphics[width=0.5\textwidth]{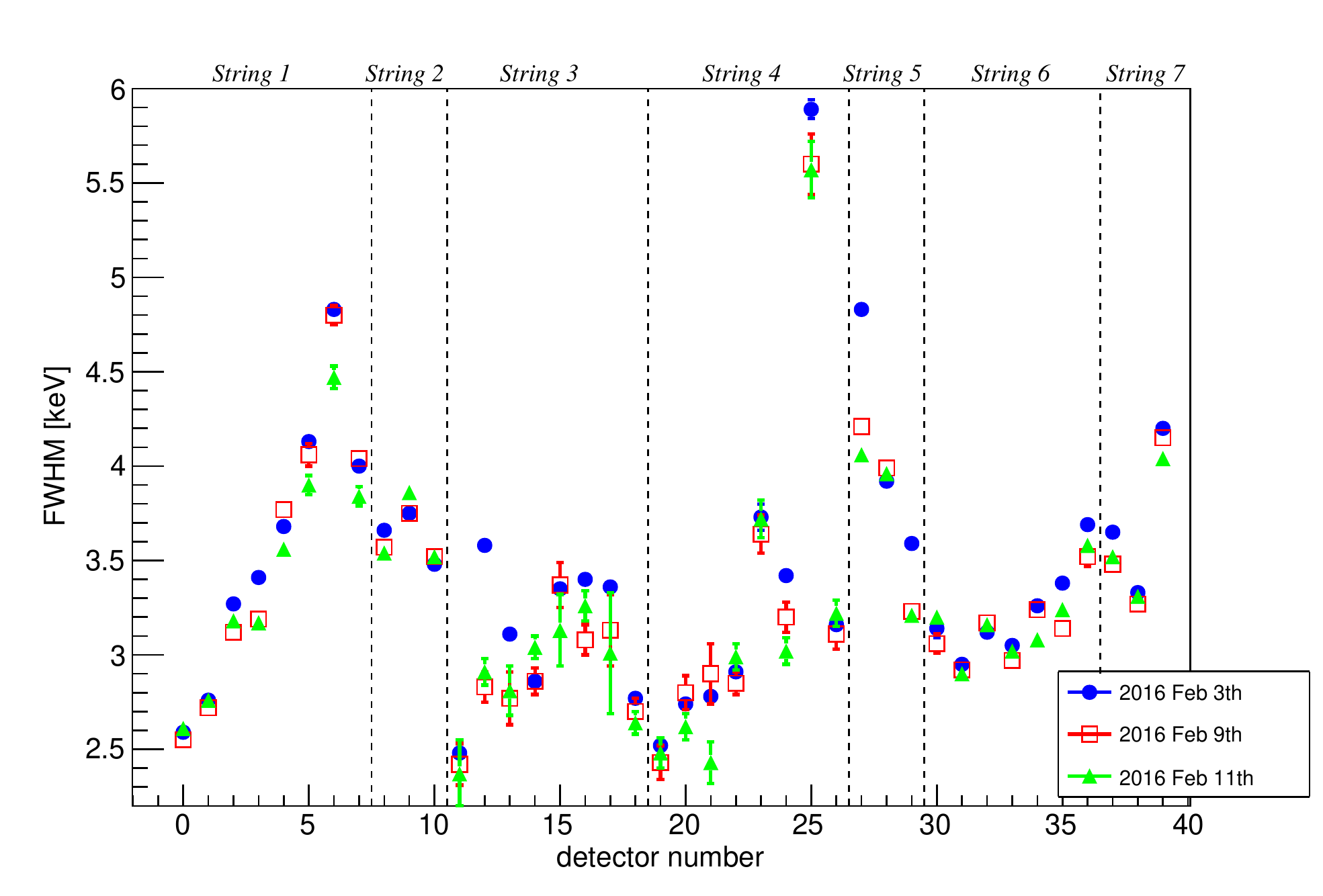}
 \caption{FWHM at $2614.5~$keV for three calibration runs during the beginning of Phase II, are included all BEGe and coaxial detectors.}
 \label{fwhm}
\end{figure}

The FWHMs of the $2614.5~$keV peak in three calibration runs of Phase II are showed in Fig. \ref{fwhm}; both for BEGes and coaxials, in the figure the horizontal axis follows the detector progressive number and the top legend reports the string number.
Strings 1, 3 and 4 contain only BEGes, string 2, 5 and 7 only coaxials, string 6 contains six BEGes and one coaxial.
All the values reported are calculated using the optimal ZAC filter for each detector.

\begin{figure}[!t]
 \centering
 \includegraphics[width=0.5\textwidth]{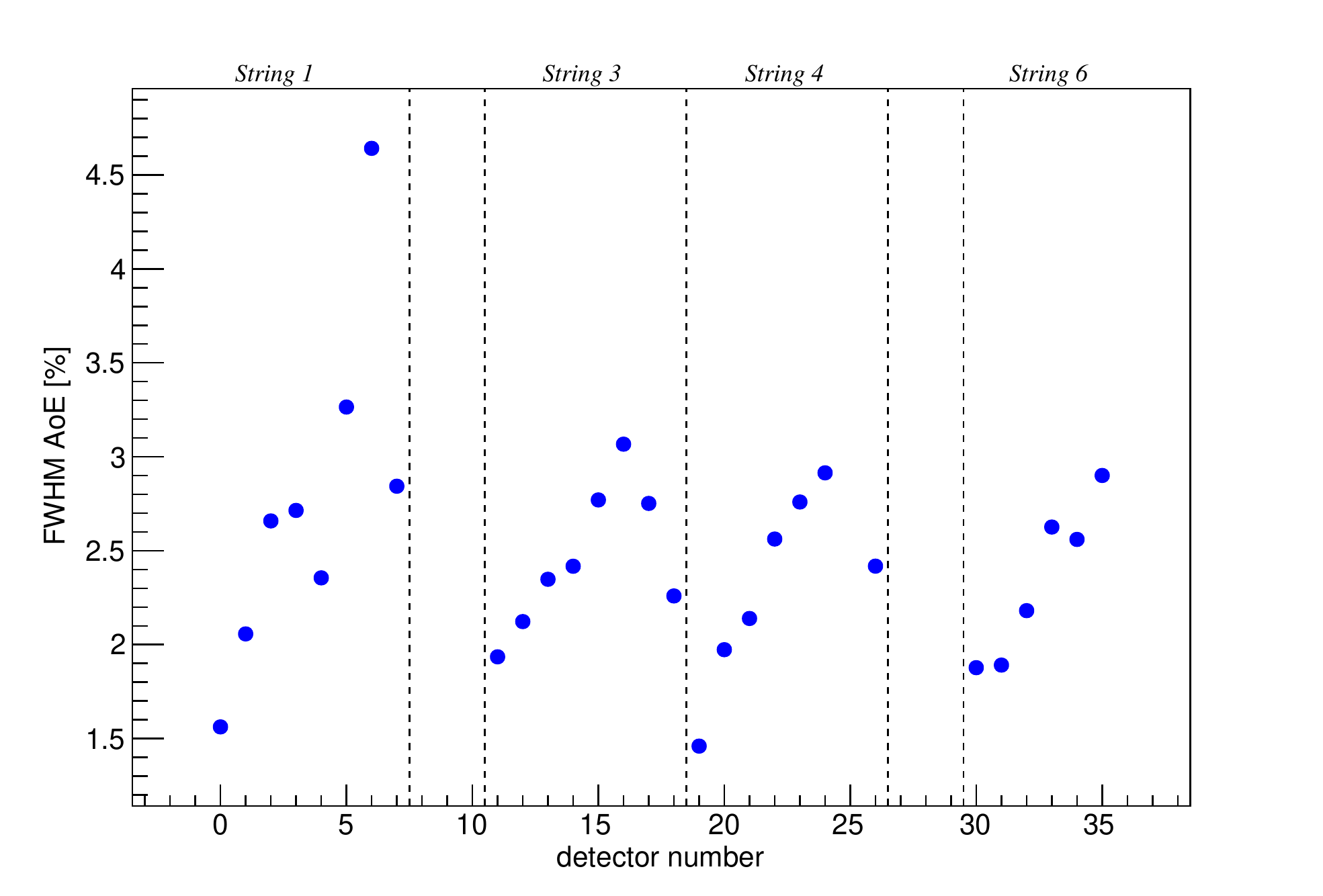}
 \caption{FWHM of the $A/E$ ratio for DEP events in BEGe detectors at the beginning of Phase II data taking}
 \label{fwhm_aoe}
\end{figure}
\begin{figure}[!t]
 \centering
 \includegraphics[width=0.5\textwidth]{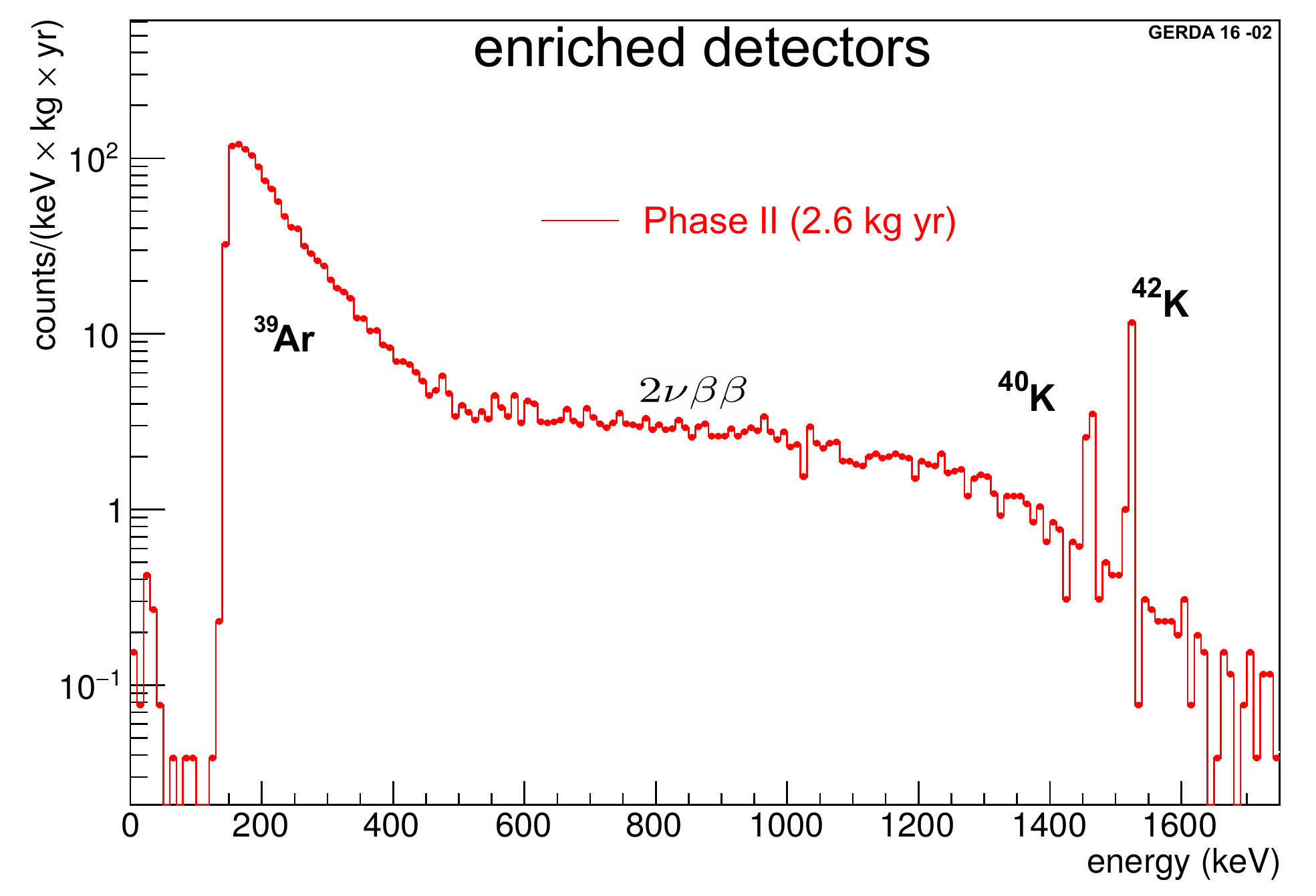}
 \caption{Background spectrum of beginning of Phase~II}
 \label{fullSpectrum}
\end{figure}

The Phase II energy resolutions, evaluated for the three considered calibrations, is in the range from $2.4~$keV to about $5.9~$keV  for the BEGes and from $3.2~$keV to $4.9~$keV for the coaxials.
If the detectors with known problems (number 6, 25 and 27) are not included, the FWHM ranges are $2.4~$keV - $4.0~$keV and $3.2~$keV~-~$4.1~$keV for the BEGes and the coaxials respectively.

Fig. \ref{fwhm} reveals that the BEGes  energy resolution correlates to the detector position in the string, that in turn scales with the length of the detector to JFET contact. This trend is found for all the four BEGe strings and  is still not completely understood, is not fully unexpected: the longer the contact, the larger its stray capacitance. This is confirmed by the analysis of the FFT noise spectra, not shown here: the white noise scales with the detector-to-JFET distance. Detailed analysis is in progress to model and finally improve the noise.
For the coaxials the same trend is not found: apart from individual detector peculiarities, this may be related to the different ratios of parasitic vs intrinsic detector capacity for the BEGe than for the coaxial detectors. The first have an intrinsic capacity of a couple  of pF, while the latter have tens of pF: hence the stray capacitance of long contacts and the parasitic of the exposed BEGe p$^+$ contacts, accounts for a larger fraction of their intrinsic capacity. This hypothesis has to be verified.

Another important parameter to evaluate the performances of the BEGe detectors is the resolution of the $A/E$ ratio; this is evaluated on the double escape peak (DEP) of the $^{208}$Tl $2614.5~$keV line. This line shows up at $1592.5~$keV in the calibration spectra; because of the specificity of the $\gamma$-pair production interaction followed by the double escape of the two $511~$keV \gammas~ generated in the e$^+$ annihilation, this line is mostly populated by SSE events. The resolution of the $A/E$ parameter distribution of DEP events is an indicator of the capability of resolving SSE from MSE, hence to discriminate the background.

Fig. \ref{fwhm_aoe} shows the percentage FWHM of $A/E$ for BEGe detectors combining the first $^{228}$Th Phase II calibration runs.
It shows the same trend of Fig. \ref{fwhm}: the FWHM of the $A/E$ ratio also scales with the position of the detector in the string and better values (around $1.5~\%$) are obtained for the top detectors.

A first Phase II background spectrum cut at $1.7~$MeV, normalized to the present exposure of $2.6~$kg$\cdot$yr, is shown in Fig. \ref{fullSpectrum}.

The spectrum shows the expected prominent structures: the low energy region (up to $500~$keV) is dominated by the long-lived $^{39}$Ar isotope; from $600$ to $1400~$keV the $2\nu\beta\beta$ spectrum shows up; then the $1462~$keV and $1525~$keV $\gamma$-lines from $^{40}$K and $^{42}$K respectively are visible.


\section{Conclusion}
After \textsc{Gerda} Phase I established a new important limit on $0\nu\beta\beta$ decay of $^{76}$Ge, the Phase II started in January 2016.

All detectors are working and the performances of the first period of data taking indicate that the noise could be improved mitigating the parasitic sources; this will improve both energy resolution and PSD performances.

The \textsc{Gerda} Phase II physics program is to take data for about three years in order to obtain $100~$kg$\cdot$yr of exposure and searching for the \gesix~ \onbb~ or improving both the half-life and the effective Majorana neutrino mass limits, possibly attaining the top border of the neutrino masses inverted hierarchy region. 

\section*{Acknowledgment}
The \textsc{Gerda} experiment is supported financially by the German Federal Ministry for Education and Research (BMBF), the German Research Foundation (DFG) via the Excellence Cluster Universe, the Italian Istituto Nazionale di Fisica Nucleare (INFN), the Max Planck Society (MPG), the Polish National Science Center (NCN), the Foundation for Polish Science (MPD programme), the Russian Foundation for Basic Research (RFBR), and the Swiss National Science Foundation (SNF). The institutions acknowledge also internal financial support. The \textsc{Gerda} collaboration thanks the directors and the staff of the LNGS for their continuous strong support of the \textsc{Gerda} experiment.

\ifCLASSOPTIONcaptionsoff
  \newpage
\fi


\begin{thebibliography}{1}
\bibitem{gerda_ep-2013}
K. H. Ackermann et al., \textsc{Gerda} Collaboration, Eur. Phys. J. C \textbf{73}, 2330 (2013) \href{http://arxiv.org/abs/1212.4067}{arXiv:1212.4067.}
\bibitem{sensitivity}
F.T. Avignone III, G.S. King III, Y.G. Zdesenko, New J. Phys. \textbf{7}, 6 (2005).
\bibitem{hdm}
M. Gunther et al., Phys. Rev. D \textbf{55}, 54 (1997).
\bibitem{igex}
C.E. Aalseth et al., Nucl. Phys. Proc. Suppl. \textbf{48}, 223 (1996).
\bibitem{bege}
M. Agostini et al., \textsc{Gerda} Collaboration, Eur. Phys. J. C \textbf{75}, 39 (2015) \href{http://arxiv.org/abs/1410.0853}{arXiv:1410.0853}
\bibitem{gtf}
L. Baudis et al., Phys. Rep. \textbf{307}, 301 (1998).
\bibitem{gtf2}
H.V. Klapdor-Klingrothaus et al., Nucl. Instrum. Methods A \textbf{481}, 149 (2002).
\bibitem{bkg}
M. Agostini et al., \textsc{Gerda} Collaboration, Eur. Phys. J. C \textbf{74} 2764 (2014) \href{http://arxiv.org/abs/1306.5084}{arXiv:1306.5084.}
\bibitem{gerdaI-res}
M. Agostini et al., \textsc{Gerda} Collaboration, Phys. Rev. Lett. \textbf{111}, 122503 (2013) \href{http://arxiv.org/abs/1307.4720}{arXiv:1307.4720.}
\bibitem{klapdor}
H. V. Klapdor-Kleingrothaus et al., Phys. Lett. B \textbf{586}, 198 (2004) \href{http://arxiv.org/abs/hep-ph/0404088}{arXiv:hep-ph/0404088.}
\bibitem{2nubb-majoron}
M. Agostini et al., \textsc{Gerda} Collaboration, Eur. Phys. J. C \textbf{75}, 416 (2015) \href{http://arxiv.org/abs/1501.02345}{arXiv:1501.02345.}
\bibitem{2nubb_excited}
M. Agostini et al., \textsc{Gerda} Collaboration, J. Phys. G: Nucl. Part. Phys. \textbf{42} (2015) 115201 \href{http://arxiv.org/abs/1506.03120}{arXiv:1506.03120.}
\bibitem{canberra}
D. Budj\'a\v{s} et al., J. Instrum. \textbf{4}, P10007 (2009).
\bibitem{cc3}
S. Riboldi et al., (NSS/MIC), (2012) IEEE, p. 782-785.
\bibitem{pmt}
M. Agostini et al., \textsc{Gerda} Collaboration, J. Phys.: Conf. Ser. \textbf{375}, 042009 (2012).
\bibitem{sipm}
J. Janicsk\'o Csáthy et al., Nucl. Instr. Methods \textbf{654}, 225-232 (2011).
\bibitem{zac}
M. Agostini et al., \textsc{Gerda} Collaboration, Eur. Phys. J. C \textbf{75}, 255 (2015) \href{http://arxiv.org/abs/1502.04392}{arXiv:1502.04392.}
\bibitem{Gatti-Manfredi} 
E. Gatti, P. F. Manfredi, Riv. Nuovo Cim. 9, 1 (1986).

\end{thebibliography}
\end{document}